\documentclass[11pt,twoside]{article}
\usepackage{asp2010}
\resetcounters

\begin{document}

\title{Observations of binaries  in AGB, post-AGB stars and Planetary Nebulae}
\author{Eric Lagadec$^1$, Olivier Chesneau$^1$
\affil{$^1$Laboratoire Lagrange, UMR7293, Universit\'e de Nice Sophia-Antipolis, CNRS, Observatoire de la C\^ote d'Azur, Boulevard de l'Observatoire, 06304 Nice Cedex 4, France}}

\begin{abstract}
During the last years, many observational studies have revealed that
binaries play an active role in the shaping of  non spherical
planetary nebulae. We review the different works that lead to the
direct or indirect evidence for the presence of binary companions
during the Asymptotic Giant Branch, proto-Planetary Nebula
and Planetary Nebula phases. We  also discuss how these binaries can influence the
stellar evolution and possible future directions in the field.
\end{abstract}

\section{Introduction}
While most of the low and intermediate mass stars (between $\sim$ 0.8
and  $\sim$ 8 M$_{\odot}$) appear more ore less spherical on the main
sequence or on the Red Giant branches , Planetary Nebulae (PNe) can harbour a wide variety of shapes and be elliptical, bipolar or multipolar.
During the last years, more and more evidences have been gathered that this departure from spherical  was linked to the influence of a binary companion. In these paper, we
will review the different works that lead to the discovery of direct
and indirect evidences for binary companion in the heart of planetary
nebulae, post-AGBs and AGB stars and discuss possibilities to detect
more binaries in AGB stars.
\section{Equatorial overdensities and jets}
The shaping of planetary nebulae  has been the subject of studies for
decades now. As early as in the late 1970s, Kwok et al. (1978)
proposed that the thin shells of planetary nebulae were due to the
interaction of a fast stellar wind (the PN wind) with the slower,
denser material ejected during the AGB phase. This Interacting Stellar
Wind Model (ISW) was very efficient at explaining the observed morphology
and density of PNe shells. During the next decade,  observation of PNe
revealed that  collimated outflows were common in PNe (Balick et al.,
1987). It was suggested that equatorial overdensities could lead to
the formation of such outflows. Many models have since been proposed to
explain the formation of the central overdensities and bipolar jets,
including the presence of a binary companion (via e.g. common envelope
evolution or Roche-lobe overflow) or magnetic fields. A very
interesting review on this topic was written by Balick and Frank
(2002). High angular resolution and high contrast observations of PNe
with the {\it Hubble
Space Telescope} (HST) revealed a vast variety of shapes and provided us
images widely used for outreach. These images also confirmed that
aspherical PNe were common and that spherical PNe were more the
exception than the rule. Dark lanes observed in the optical images
were clear indirect proofs of the presence of equatorial dusty
structures (see e.g. Matsuura et al., 2005). Unfortunately,
observations in the optical enable us to study those equatorial
structures via scattered light only. Observations at longer wavelengths are
needed to directly characterise the
equatorial overdensities needed to explain the formation of bipolar
PNe. This was made possible with the advance  of high angular
resolution techniques in the infrared (e.g. adaptive optics and
interferometry) and the millimetre (imaging and interferometry). Infrared
observations, using interferometry (e.g., Chesneau et al., 2006, 2007,
Lykou et al., 2011) or direct imaging with adaptive optics (e.g
Lagadec et al., 2006), resolved these equatorial structures. It was
also achieved in the millimetre domain with CO observations (e.g. Peretto
et al., 2007, Alcolea et al., 2007). The infrared observations are
sensitive to dust emission and help us study the dust spatial
distribution and content. The millimetre observations enable us to
study the CO gas spatial distribution and dynamics, thanks to their
spectral resolution. Two kinds of equatorial overdensities have thus
been observed: torii and stratified discs.
 
Torii are massive (masses of the order of $\sim$ a solar mass or more), have a low
expansion velocity (typically a few km/s, see e.g. Peretto et
al. (2007)). Their kinematic is dominantly radial and their angular
momentum is limited. They are short-lived, so that if the mass loss
stops, the material will rapidly expand and vanish.

Discs exhibit  clear vertical stratifications, with scale height  governed by the
gas pressure only. They have very small aperture angle (less than
$\sim$10 degrees typically)  and their kinematic is Keplerian, with a small expansion
component ($<$10 km/s) (see e.g. Bujarrabal et al., 2013, Deroo et
al., 2007). Their lifetimes are much larger than the torii
described above and are comparable or larger than the typical 
lifetime of a PN, which is typically few
tens of thousand years (van Winckel 2003).

High angular observations, mostly with the HST, also revealed the
presence of multipolar PNe. The formation of such nebulae can not be
explained with an isotropic wind interacting with an equatorial
density. Sahai \& Trauger (1998) proposed that this could be due to
precessing jets. The presence of jets was confirmed by a study of
proto-PNe by Bujarrabal et al. (2001). They used CO observations to
measure the mass, linear momentum and kinetic energy of bipolar flows
in proto-PNe. They found that in about 80\% of the PPNe, the momentum
of the outflow is too high to be powered by radiation pressure only
(up to $\sim$ 1000 times larger). An extra source of angular momentum
is thus needed to explain the presence of these jets.

\section{Binaries as shaping agents}
So far, we have shown that equatorial overdensities and jets are
shaping most of the PNe. The question that need to be answered now is
how are these discs/torii and jets formed. Different models have
been proposed to explain the formation of jets, involving either the
magnetic field  (e.g.  Garcia-Segura et al.,
2005) or the influence of a binary companion (stellar or substellar)
as the main shaping agent (e.g. Soker at al., 2004). Two key works
certainly settled the debate by considering the energy and angular
momentum carried by the magnetic fields expelled from AGB stars. In 2005, Noam Soker claimed that: ``{\it a single star
  can not supply enough energy and angular momentum to shape those
  nebulae}''. And later, in 2006, Nordhaus et al. showed that
magnetic fields can play an important role in the shaping of bipolar PNe but isolated stars can not sustain a magnetic field for long enough. Magnetic field can thus play a role to collimate jets, but the
angular momentum they need to be sustained requires the presence of a
binary companion.

\section{Direct detection of binaries in PNe}
As soon as in 2005, it was quite convincingly shown, from a  theoretical
point of view, that binary companions should be the main shaping agents
of PNe. But, by then, only a handful of binary companions were
known. Following an idea by Orsola de Marco, a  community effort
collaboration started during the Asymmetrical Planetary Nebulae IV conference held in
La Palma, aiming at hunting for binaries in the heart of PNe:
PLANB\footnote{http://www.wiyn.org/planb/}. Three main methods were
used to look for binaries in PNe. The study of flux variability can
tell us about eclipses, tidal deformations induced by the companion or
irradiation effects. Spectral variability is a measure of radial
velocities and enables the discovery of companions as it has been
widely shown in the exoplanets community. Finally, central stars of PNe
being hot, searching for infrared excess in their core can lead to the
detection of cool companions.

\begin{figure}
\begin{center}
\includegraphics[width=11cm]{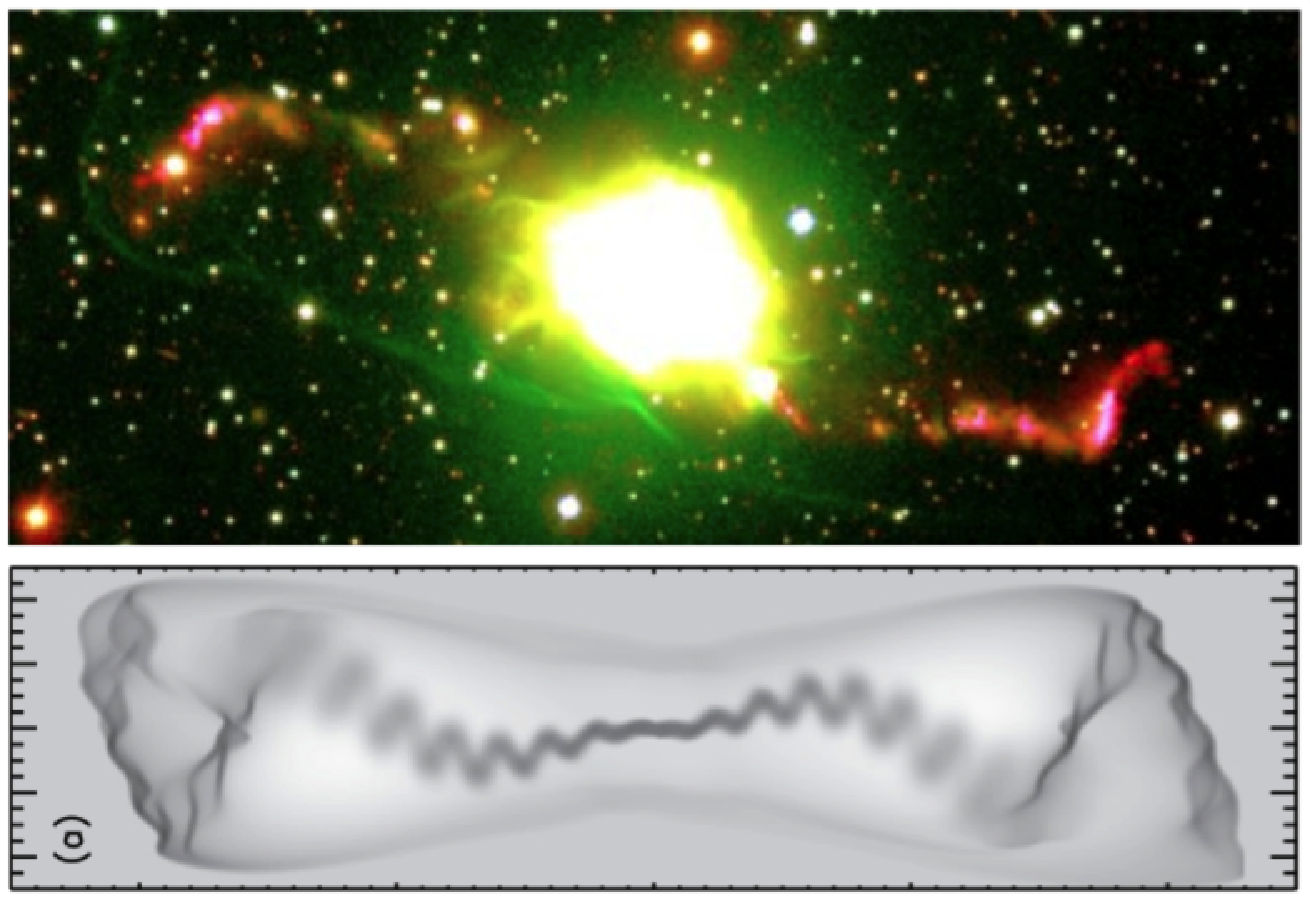} 
\caption{Figure 1: Up: FORS2/VLT colour-composite image of Fleming 1 showing the [O
  III] envelope around the bipolar jets (from Boffin et al., 2012). The field of view is 5.5’ x
2.3’. The jets shows clear indication of precession. Down:
three-dimensional gas dynamical simulation from a source in a circular
orbit, and with a precessing outflow axis  (Raga et al., 2009).}
\end{center}
\end{figure}

The first  big leap forward was made thanks to a variability study
using OGLE data (Miszalski et al., 2009a). It was then identified that
binary stars were discovered in PNe sharing common characteristics
(Miszalski et al., 2009b). These characteristics are intuitively
bipolarity, but also the presence of low ionisation filaments and
equatorial rings.  Using this, new discoveries were made using both
photometric and spectroscopic surveys (e.g. Jones
et al. (2010, 2012, 2013); Miszalski (2011a,b,c,2013)
;Boffin 2012). Spatio-kinematical models (e.g. Jones et al., 2010) were used to lift the degeneracy due to projection effects (a bipolar
nebula seen pole-on will appear circular in the sky). These works demonstrate
that the orbital plane of the
discovered binaries are coincident with the observed equatorial
overdensities and perpendicular to the bipolar/multipolar
lobes. This is a nice confirmation that binaries play an active role
in the shaping of bipolar PNe.

An interesting outcome of these studies is that most of the systems
are short period binaries (P$< 3$ days) and certainly went through a
common envelope evolution which lead to the shrinking of the orbit
(Miszalski et al., 2009).
 The observed jets appear to be
older than the nebulae. It is very likely that the jets were produced
during an interaction before a common envelope phase that lead to the
formation of the nebula. Fleming 1 (Fig.1)  provides a very nice textbook
case for this scenario (Boffin et al., 2012). Radial velocities
measurements show that its core harbour a binary system with a $\sim$
1.2 days period. The precessing jets have a timescale of 10$^4$ years,
so that the orbital period while the jets were created should have
been $\sim$10$^2$-10$^3$ years. But, as measured by radial velocities,
the orbital is much shorter. It is thus very likely that the jets
formed before a common envelope phase. The orbit shrunk during that
phase to become the present close binary system. This lead Boffin et
al to claim that: {\it  ``Similar
binary interactions are therefore likely to explain these kinds of outflows in a large
variety of systems''}.

\section{Direct detection of binaries in post-AGB stars}
As shown before, more and more binary systems are being discovered in
PNe. The detection of binaries in post-AGB systems is made more complex
by the fact that the central stars are pulsating and embedded in
dust. This makes the three techniques used for hunting binaries in PNe
(radial velocities, photometric variation and infrared excess)
difficult to apply. Morphological studies using high angular
resolution infrared images show that spherical proto-PNe (post-AGB
stars on their way to form PNe) are very rare (Lagadec et al., 2011).
O-rich PPNe are predominantly bipolar or multipolar. Their low C/O
could be due either to the interaction with a  binary companion during a
common envelope phase or hot bottom burning converting carbon to
nitrogen (De Marco, 2009). The common envelope scenario will lead to
an ejection of the envelope earlier than during single star evolution,
leading to a lower C/O ratio, as less carbon is dredged-up to the
surface (Izzard et al. 2006). Hot bottom burning occurs in the most
massive AGB stars, making it likely that the bipolar PPNe have
progenitor with larger masses than the elliptical ones in agreement with the work by Corradi \& Schwartz
(1995).  This could be explained in the frame of the binary system
progenitors paradigm  as primaries that undergo a common envelope phase, and thus become
 bipolar, tend to have a higher mass (Soker, 1998) . 

 Binaries have however been discovered in two
emblematic bipolar post-AGBs.  From the position of the different
masers in OH\,231.8+4.2, Gomez et al. (2001) found that it was a binary
system, but the central star of OH\,231.8, QX\,Pup is actually a Mira
star. A binary system was also discovered in the iconic post-AGB star
the Red Rectangle (Waelkens et al., 1996), with a period of 318
days. This orbit is too small to host an AGB star, so that the object
must have evolved from a binary channel.  It is  widely agreed
that the bipolar morphology of this object is due to the interaction
with a binary companion.

A very interesting systematic search for binaries in post-AGBs has
been performed by Bruce Hrivnak and his undergraduate students at
Valparaiso University. They have been doing radial velocities
monitoring of post-AGB stars till 1994, hunting for long period
binaries. They might have detected a binary system with P$>$22 years
(Hrivnak et al., 2011). Their result tend to indicate that potential
binary companions have e periods greater than 25 yr or masses of brown
dwarfs or super-Jupiters.

Another large scale hunt for binaries in post-AGB was initiated by Hans
van Winckel and collaborators. Their targets were selected based on
the presence of a near-infrared excess in their spectral energy
distribution  (de Ruyter et al., 2006).  This excess is due to the
presence of dust near the sublimation temperature, in a stable,
 compact (R$\sim$ 10 AU), Keplerian disc, as confirmed  by their infrared
interferometric measurements (Deroo et al., 2007). Radial velocities
monitoring indicates that close binary systems (0.5-3 AU) are present in the core of
those discs (van Winckel et al., 2009). They also found that all the
discs were oxygen-rich and no photospheric evidence for dredge-up. The
binary companion thus has an influence on the chemical evolution of
the star and can prevent dredge up from forming carbon-rich objects. Such discs are
very likely to be formed in all the binary systems too small to
accommodate a fully grown up AGB star. Compact, Keplerian discs in
post-AGB stars  are
thus very likely indirect evidence for binary interaction.
 To quote Olivier Chesneau: {\it ``My personal opinion is that the discovery of a stratified disk with proved Keplerian
kinematics is directly connected to the influence of a companion, albeit the few
exceptions presented above, namely the Young Stellar Objects or the critical velocity
rotating massive sources such as Be stars. This hypothesis must be confirmed by
further observations''.}

\section{Detection of binaries in AGB stars}
The direct detection of a companion by the aforementioned techniques
in an AGB star is made very difficult by the large scale pulsation of
AGB star and their infrared brightness due to the dust in their
envelopes.

High angular resolution observations of their envelopes have revealed
that many AGB stars are actually asymmetrical. The iconic
carbon-rich AGB star is known to display a more or less spherically symmetric
envelope at large scale, as seen through dust scattered ambient
Galactic light  (Mauron \& Huggins, 1999; Leao et al.,
2006). However, when one peers deep into its core using infrared high
angular resolution observations, such as speckle imaging (Weigelt et
al., 1998) or adaptive optics (Leao et al., 2006), one can clearly
detect the presence of clumps. CO observation of V Hya also revealed
the presence of a fast, bipolar outflow and an equatorial disc (Hirano
et al., 2004, Sahai et al., 2003). A disc was also detected around one
of the closest AGB stars, (L$_2$ Pup) using near-infrared adaptive
optics imaging (Kervella et al., 2014). The morphologies of these
objects seems to indicate that they are the progenitors of bipolar
PNe. If one accepts the proposed idea that bipolar PNe are due
to binary interaction, their progenitors should be binary systems. The
AGB stars we mentioned before are thus very likely binary systems. 

\begin{figure}
\begin{center}
\includegraphics[width=14cm]{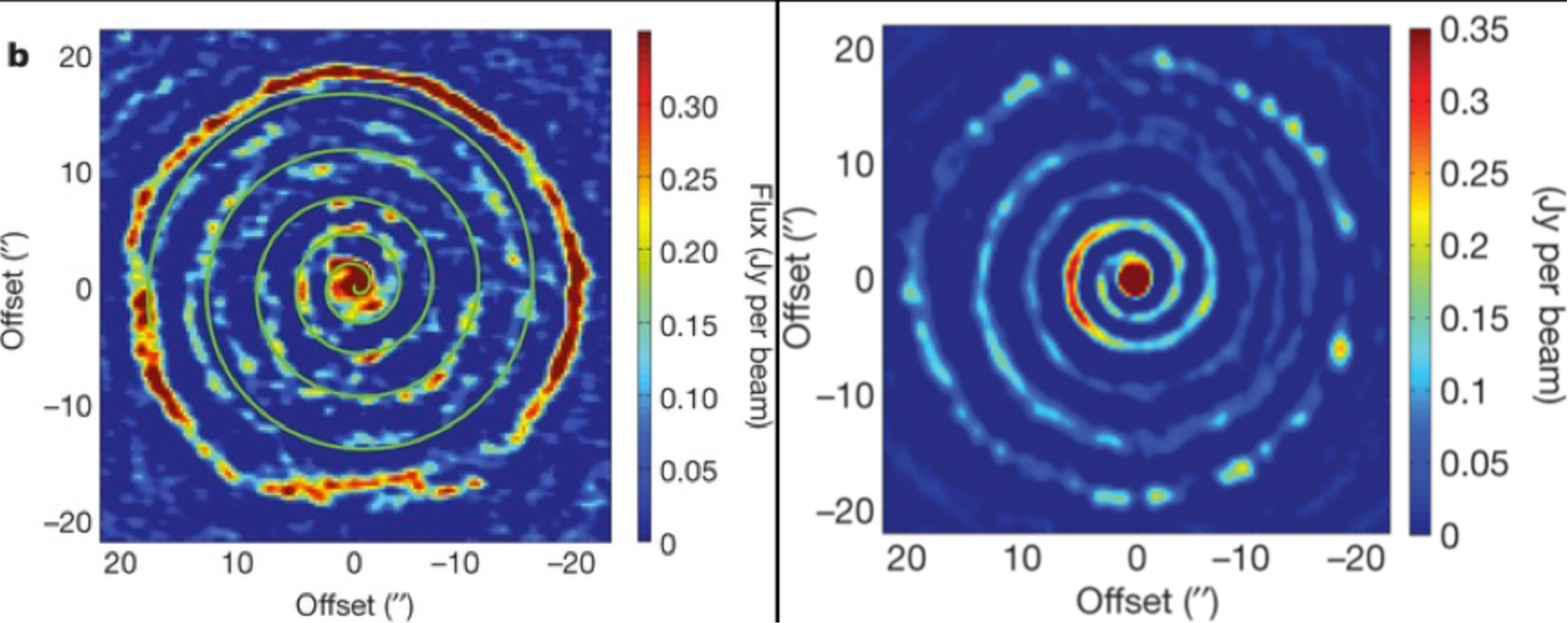} 
\caption{Figure 2: Left: ALMA CO 3-2 map of the AGB star R Scl, revealing the
  presence of a detached shell and a spiral pattern. Right:  SPH model
of an interacting binary system containing an AGB transferring material
to his companion via Wind Roche-Lobe overflow to form the 
observed spiral pattern.}
\end{center}
\end{figure}

It is very difficult to directly detect those binaries, but indirect
techniques can be used. Using hydrodynamics simulation of dusty winds
in binary systems, Mastrodemos \& Morris (1999) have shown that shocks
between the wind of an AGB star and a companion star can lead to the
formation of a dusty spiral covering most of the solid angle around
the binary. Mohamed and Podsiadlowski (2007) found that a similar
spiral could be the outcome of the interaction an AGB wind with a binary with
simulation of a new mass-transfer mode: wind Roche-lobe overflow. When
the wind acceleration occurs at a few stellar radii, close to the
Roche-lobe, as in Mira stars, the wind can fill the Roche
lobe. Material can then be transferred, like in traditional Roche-lobe
overflow. One of the most striking results obtained with ALMA so far
is the direct detection of such a spiral pattern around the AGB star
R Scl (Fig.2, Maercker et al., 2012). Detecting spiral around AGB star
is thus an indirect evidence for a binary companion. Those spirals are
Archimedes spiral with a constant spacing. Knowing the pitch of the
spiral and the expansion velocity of the wind (which can be measured
with ALMA), one can easily determine the period of the binary system.
Such spiral patterns have been detected around other AGB stars in
reflected light for AFGL 3068 (Mauron \& Huggins (2006) and CIT 6 (Kim
et al., 2013) and with ALMA CO observation in Mira (Ramstedt et al.,
2014) and for IRC +10216
(Homan, these proceedings) . One way to find a large number of interacting binaries in AGB
stars would be to perform surveys looking for such spiral
patterns. Such surveys could be done either with large scale optical
imagers, looking for dust reflection by ambiant Galactic light, high
angular resolution adaptive imaging (probably with coronography to
reach high contrasts and why not directly detect the companion) or direct CO imaging with millimetre
interferometers such as ALMA. The two first techniques probe the dust,
while CO observations probe the gas and also measure the shell
expansion velocities, leading to an accurate determination of the
binary separation. Another interesting approach is to perform a
multi-scale/multi-wavelengths approach and study the circumstellar
material from the dust formation radius till its interaction with the
interstellar medium (see Mayer and Paladini, these proceedings).

Another technique to indirectly discover binaries in AGB star is
looking for ultraviolet excess. AGB stars being
cool, they are very faint in the UV. Sahai et al. (2008) performed a
small UV imaging survey using GALEX). They observed 25 AGB stars and
found a UV excess in 9 of them. This UV excess likely result either
from the presence of a hot companion, or from accretion induced by a
companion.

\begin{figure}
\begin{center}
\includegraphics[width=8cm]{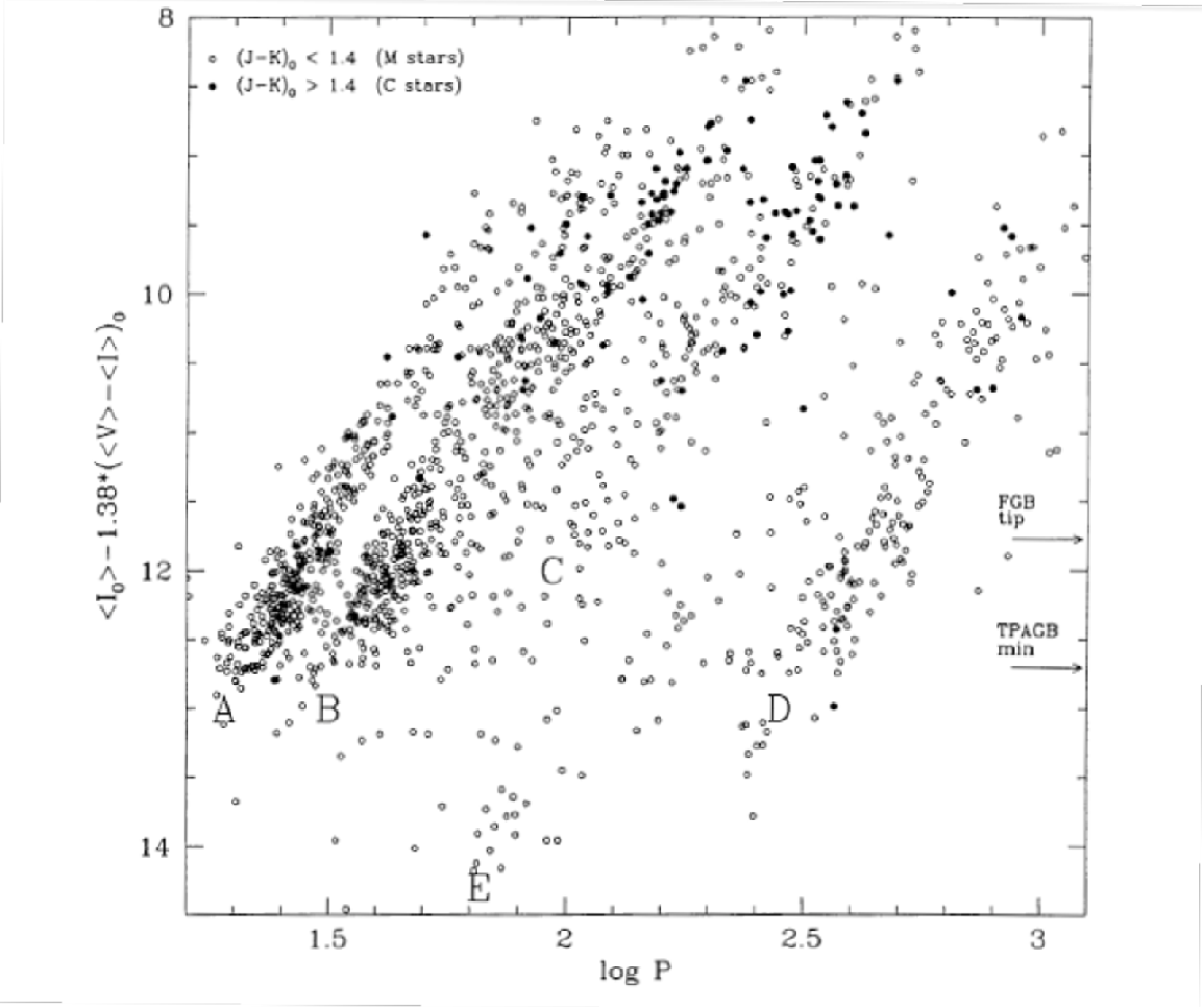} 
\caption{Figure 3:Period luminosity diagram for red giants in the Large
  Magellanic Cloud ($<$I$>$ and $<$V$>$ are mean magnitudes and P i in days.
  Star on sequence A to D are giant stars pulsating in the
fundamental modes and different overtones. The stars on sequence E
have a light variability due to the presence of a binary
companion. (Figure from Wood (1999))}
\end{center}
\end{figure}

Finally, binaries in AGB stars can be found using observations in
nearby galaxies. As we know the distance to those galaxies,
luminosities can be accurately determined, unlike in the Milky Way. Determining period for the
AGB variables enable the  obtention of period luminosity diagrams
(Fig. 3; see
e.g. Wood et al., 1999 for the Large Magellanic Cloud (LMC)).  Five
distinct period-luminosity sequences have been found on the low mass
giant branch. Star on sequence A to D are giant stars pulsating in the
fundamental modes and different overtones. The stars on sequence E
have a light variability due to the presence of a binary companion
(see  e.g. Nicholls et al., 2010). Those stars are likely the
precursors of PNe with close binary central stars and binary post-AGB
and post-RGB stars.

These stars are likely the immediate precursors of planetary nebulae (PNe)
with close binary central stars as well as other binary
post-asymptotic giant branch (post-AGB) and binary post-red giant
branch (post-RGB) stars. Binary post-RGB stars are similar to binary
post-AGB stars, but the interaction with the companion lead to the
formation of a disc as soon as the RGB phase. It is almost impossible
to distinguish post-RGBs from post-AGBs in the Galaxy as the
main difference between the two classes of objects is their
luminosities. Post-RGB stars can be more easily discovered in nearby
galaxies with known distances, where there luminosity can be easily
determined (see e.g. Kamath et al., 2014).  A population study of
binary AGBs, post-RGBs and post-AGBs in the LMC show that between 3
and 19 \% of the PNe come from single stars while $\sim$ 50\% of the PNe
with close binary precursors have post-RGB precursors (Nie et al.,
2012).  The interaction with a binary companion can thus prevent a low
or intermediate mass star from becoming an AGB star.

\section{Conclusions and perpectives}
In this review, we have shown that binarity was common in AGB,
post-AGB (post-RGB)  stars and PNe. If the binary companion is
close/large enough, it can lead to the formation of a common envelope
or a disc. The presence of a binary can affect the chemical evolution
of a star by quickly ejecting its circumstellar envelope. This can
prevent stars from becoming carbon-rich, as the envelope is ejected before
dredge-up brings enough carbon to the surface to make it
carbon-rich. If the companion is close enough, this can even prevent
the star from becoming an AGB star and form a post-RGB star. In case of
mass transfer, the secondary can be enriched in s-processed elements
and lead to the formation of peculiar objects such as the Barium
stars, the carbon-enhanced metal poor stars or the CH stars (see the nice review by Jorissen, 1999).

To better understand the impact of binarity on the evolution of low
and intermediate mass stars, we need to better understand how binaries
form discs and how these discs evolve. These discs will have an impact
on the chemical composition of the stars, as dust will form and
refractory elements will deplete on dust. These discs could also impact
the evolution timescales of the central stars and change their
mass-loss history and thus their chemical enrichment. 

Finally, to better understand how binarity affects the evolution of a
star during the AGB, we should start a large scale survey of AGB stars
with binary companion, to characterise its population (binary
fraction, mass ratio and
separation of the components) and study how these parameters affect the chemical evolution of the
AGB stars. This will enable to quantify the impact of binarity on the
chemical evolution of galaxies and maybe start a new series of
conferences: {\it Why do galaxies care about binary AGB stars?''}.

\acknowledgements This review is dedicated to the memory of Olivier
Chesneau, who left us too early. Olivier was a bright and passionate
astronomer who inspired many young researchers and opened new horizons
to the evolved star community. E.L. would like to thanks David Jones,
Henri Boffin, Hans van Winckel and Albert Zijlstra for fruitful
discussions during the preparation of this review.

\bibliography{aspauthor}
Alcolea, J., Neri, R., \& Bujarrabal, V.\ 2007, \aap, 468, L41 \\
Balick, B., Preston, 
H.~L., \& Icke, V.\ 1987, \aj, 94, 1641 \\
Balick, B., \& Frank, A.\ 2002, \araa, 40, 439 \\
 Boffin, H.~M.~J., Miszalski, B., Rauch, T., et al.\ 2012, Science, 338, 773 \\
Bujarrabal, V., Castro-Carrizo, A., Alcolea, J., \& S{\'a}nchez Contreras, C.\ 2001, \aap, 377, 868 \\
Bujarrabal, V., Castro-Carrizo, A., Alcolea, J., et al.\ 2013, \aap,
557, L11\\
Castro-Carrizo, A., Neri, R., Bujarrabal, V., et al.\ 2012, \aap, 545,
A1  \\
Corradi, R.~L.~M., \& Schwarz, H.~E.\ 1995, \aap, 293, 871 \\
Chesneau, O., Lykou, F., Balick, B., et al.\ 2007, \aap, 473, L29 \\
Chesneau, O., Collioud, A., De Marco, O., et al.\ 2006, \aap, 455, 1009 \\
Deroo, P., Acke, B., Verhoelst, T., et al.\ 2007, \aap, 474, L45 \\
De Marco, O.\ 2009, \pasp, 
121, 316 \\
de Ruyter, S., van Winckel, H., Maas, T., et al.\ 2006, \aap, 448, 641 \\
Garc{\'{\i}}a-Segura, G., L{\'o}pez, J.~A., 
\& Franco, J.\ 2005, \apj, 618, 919 \\
G{\'o}mez, Y., \& Rodr{\'{\i}}guez, L.~F.\ 2001, \apjl, 557, L109 \\
Hirano, N., Shinnaga, H., Dinh-V-Trung, et al.\ 2004, \apjl, 616, L43
\\
Hrivnak, B.~J., Lu, W., Bohlender, D., et al.\ 2011, \apj, 734, 25 \\
Izzard, R.~G., Dray, L.~M., Karakas, A.~I., Lugaro, M., \& Tout, C.~A.\ 2006, \aap, 460, 565 \\
Jones, D., Boffin, H.~M.~J., Miszalski, B., et al.\ 2014, \aap, 562, A89 \\
Jones, D., Mitchell,  D.~L., Lloyd, M., et al.\ 2012, \mnras, 420, 2271 \\
Jones, D., Lloyd, M., Santander-Garc{\'{\i}}a, M., et al.\ 2010,
\mnras, 408, 2312 \\
Jorissen, A.\ 1999, 
Asymptotic Giant Branch Stars, 191, 437 \\
Kamath, D., Wood, P.~R., \& Van Winckel, H.\ 2014, \mnras, 439, 2211\\
Kervella, P., Montarg{\`e}s, M., Ridgway, S.~T., et al.\ 2014, \aap,
564, A88 \\
Kim, H., Hsieh, I.-T., Liu, S.-Y., \& Taam, R.~E.\ 2013, \apj, 776, 86
\\
Kwok, S., Purton, C.~R., 
\& Fitzgerald, P.~M.\ 1978, \apjl, 219, L125 \\
Lagadec, E., Chesneau, O., Matsuura, M., et al.\ 2006, \aap, 448, 203
\\
Lagadec, E., Verhoelst, 
T., M{\'e}karnia, D., et al.\ 2011, \mnras, 417, 32 
Le{\~a}o, I.~C., de Laverny, P., M{\'e}karnia, D., de Medeiros, J.~R., \& Vandame, B.\ 2006, \aap, 455, 187 \\
Lykou, F., Chesneau, O., Zijlstra, A.~A., et al.\ 2011, \aap, 527,
A105 \\
Maercker, M., Mohamed, 
S., Vlemmings, W.~H.~T., et al.\ 2012, \nat, 490, 232 \\
Mauron, N., \& Huggins, P.~J.\ 1999, \aap, 349, 203 \\
Mauron, N., \& Huggins, P.~J.\ 2006, \aap, 452, 257 \\
Mastrodemos, N., \& Morris, M.\ 1999, \apj, 523, 357\\
Matsuura, M., Zijlstra, A.~A., Molster, F.~J., et al.\ 2005, \mnras, 359, 383 \\
Mohamed, S., \& Podsiadlowski, P.\ 2007, 15th European Workshop on White Dwarfs, 372, 397 \\
Miszalski, B., Boffin, H.~M.~J., \& Corradi, R.~L.~M.\ 2013, \mnras, 428, L39 \\
Miszalski, B., Jones, D., Rodr{\'{\i}}guez-Gil, P., et al.\ 2011a, \aap, 531, A158 \\
Miszalski, B., Corradi, R.~L.~M., Boffin, H.~M.~J., et al.\ 2011b, \mnras, 413, 1264 \\
Miszalski, B., Miko{\l}ajewska, J., K{\"o}ppen, J., et al.\ 2011c, \aap, 528, A39 \\
Miszalski, B., Acker, A., Parker, Q.~A., \& Moffat, A.~F.~J.\ 2009b,
\aap, 505, 249 \\
Miszalski, B., Acker, A., Moffat, A.~F.~J., Parker, Q.~A., \& Udalski, A.\ 2009a, \aap, 496, 813 \\
Miszalski, B., Acker, A., Moffat, A.~F.~J., Parker, Q.~A., \& Udalski,
A.\ 2008, \aap, 488, L79 \\
Nicholls, C.~P., Wood, 
P.~R., \& Cioni, M.-R.~L.\ 2010, \mnras, 405, 1770 \\
Nie, J.~D., Wood, P.~R., 
\& Nicholls, C.~P.\ 2012, \mnras, 423, 2764\\
Nordhaus, J., \& Blackman, E.~G.\ 2006, \mnras, 370, 2004 \\
Peretto, N., Fuller, G., Zijlstra, A., \& Patel, N.\ 2007, \aap, 473,
207 \\
Raga, A.~C., Esquivel, A., 
Vel{\'a}zquez, P.~F., et al.\ 2009, \apjl, 707, L6 \\
Ramstedt, S., Mohamed, 
S., Vlemmings, W.~H.~T., et al.\ 2014, arXiv:1410.1529 \\
Sahai, R., \& Trauger, J.~T.\ 1998, \aj, 116, 1357 \\
Sahai, R., Morris, M., Knapp, G.~R., Young, K., \& Barnbaum, C.\ 2003,
\nat, 426, 261\\ 
Sahai, R., Findeisen, K., 
Gil de Paz, A., \& S{\'a}nchez Contreras, C.\ 2008, \apj, 689, 1274 \\
Soker, N.\ 1998, \apj, 496, 833\\
Soker, N.\ 2004, Asymmetrical 
Planetary Nebulae III: Winds, Structure and the Thunderbird, 313, 562 
Soker, N.\ 2006, \pasp, 118, 260 \\
van Winckel, H.\ 2003, \araa, 41, 391 \\
van Winckel, H., Lloyd Evans, T., Briquet, M., et al.\ 2009, \aap,
505, 1221\\
Waelkens, C., Van Winckel, H., Waters, L.~B.~F.~M., \& Bakker, E.~J.\
1996, \aap, 314, L17 \\
Weigelt, G., Balega, Y., Bloecker, T., et al.\ 1998, \aap, 333, L51 \\
Wood, P.~R., Alcock, C.,  Allsman, R.~A., et al.\ 1999, Asymptotic Giant Branch Stars, 191, 151 \\

\end{document}